\documentclass[preprint,12pt]{elsarticle}

\usepackage{amsmath}
\usepackage{amssymb}
\usepackage{caption}
\usepackage{graphicx}
\usepackage[version=4]{mhchem}
\usepackage{geometry}
\usepackage{setspace}
\usepackage{booktabs}
\usepackage{multirow}

\usepackage[colorlinks, linkcolor=blue, anchorcolor=blue, citecolor=blue]{hyperref}

\journal{Combustion and Flame}

\begin{document}

\begin{frontmatter}



\title{A data-driven sparse learning approach to reduce chemical reaction mechanisms}


\author[label1]{Shen Fang}
\author[label1]{Siyi Zhang}
\author[label1]{Zeyu Li}
\author[label1,label2]{Qingfei Fu}
\author[label3,label4]{Chong-Wen Zhou}
\author[label1,label2]{Wang Han}
\author[label1,label2]{Lijun Yang \corref{cor1}}

\affiliation[label1]{
    organization={School of Astronautics},
    addressline={Beihang University, Beijing 102206},
    country={PR China}}
\affiliation[label2]{
    organization={Ningbo Institute of Technology},
    addressline={Beihang University, Ningbo 315800}, 
    country={PR China}}
\affiliation[label3]{
    organization={School of Energy and Power Engineering},
    addressline={Beihang University, Beijing 102206}, 
    country={PR China}}
\affiliation[label4]{
    organization={Combustion Chemistry Centre, School of Biological and Chemical Sciences},
    addressline={University of Galway, Galway H91 TK33},
    country={Ireland}}
\cortext[cor1]{Corresponding author}

\begin{abstract}
Reduction of detailed chemical reaction mechanisms is one of the key methods for mitigating the computational cost of reactive flow simulations. Exploitation of species and elementary reaction sparsity ensures the compactness of the reduced mechanisms. In this work, we propose a novel sparse statistical learning approach for chemical reaction mechanism reduction. Specifically, the reduced mechanism is learned to explicitly reproduce the dynamical evolution of detailed chemical kinetics, while constraining on the sparsity of the reduced reactions at the same time. Compact reduced mechanisms are be achieved as the collection of species that participate in the identified important reactions. We validate our approach by reducing oxidation mechanisms for $n$-heptane (194 species) and 1,3-butadiene (581 species). The results demonstrate that the reduced mechanisms show accurate predictions for the ignition delay times, laminar flame speeds, species mole fraction profiles and turbulence-chemistry interactions across a wide range of operating conditions. Comparative analysis with directed relation graph (DRG)-based methods and the state-of-the-art (SOTA) methods reveals that our sparse learning approach produces reduced mechanisms with fewer species while maintaining the same error limits. The advantages are particularly evident for detailed mechanisms with a larger number of species and reactions. The sparse learning strategy shows significant potential in achieving more substantial reductions in complex chemical reaction mechanisms.
\end{abstract}


\end{frontmatter}


\section*{Novelty and Significance Statement}
The novelty of this research lies in proposing a novel sparse statistical learning approach for detailed chemical reaction mechanism reduction. The method extensively explores reaction sparsity by identifying dominant reactions in the mechanism with a weight criteria statistically learnt from sparse learning, ensuring highly compact mechanism reductions. Comparisons with existing methods validated the performance of reduced mechanisms from the sparse learning strategy in predicting both fundamental and turbulent combustion characteristics. This approach surpasses DRGEP by producing reduced mechanisms with fewer species under the same error limit. Significantly, in comparison with state-of-the-art methods, it achieves more thorough reductions with comparable maximum relative errors. Particularly notable is its efficacy in reducing larger mechanisms. This research provides a new way for achieving substantial mechanism reduction, thus enhancing the efficiency of reactive flow computational simulations.

\section*{Author Contributions}
\begin{itemize}
\item S. Fang: Conceptualization, methodology, software, validation, formal analysis, writing - original draft, visualization.
\item S. Zhang: Methodology, writing - review \& editing.
\item Z. Li: Methodology, conceptualization.
\item Q. Fu: Supervision, funding acquisition.
\item C.-W. Zhou: Conceptualization, validation, writing - review \& editing, supervision, funding acquisition.
\item W. Han: Writing - Conceptualization, validation, writing - review \& editing, supervision, funding acquisition.
\item L. Yang: Supervision, funding acquisition.
\end{itemize}

\newpage

\section{Introduction\label{sec:introduction}}
Reactive flow computational simulations based on detailed chemical reaction mechanisms are often hindered by the large number and stiffness of equations that need to be solved. Mechanism reduction, which aims to eliminate species and reactions considered negligible in the detailed mechanism\cite{lu09}, has been one of the most important methods for accelerating combustion reactive flow simulations. This strategy leverages the intrinsic low dimensionality or sparsity of variables within the dynamic system described by the detailed reaction mechanism\cite{mass92}. Fundamentally, only a limited number of variables (species or elementary reactions in the detailed mechanism) play dominant roles in the chemical kinetics, allowing for the removal of surplus variables while maintaining prediction accuracy\cite{correa00}. The exploitation of this sparsity has driven the development of numerous methodologies for mechanism reduction\cite{tomlin95, lu06}.

Directed relation graph (DRG)-based methods, proposed by Lu et al.\cite{lu05}, explore species sparsity by discerning species' contributions to crucial reaction fluxes. The simplicity and reliability of DRG-based methods have made them prominent in mechanism reduction. Many advanced reduction methods have emerged from the backbone DRG method, such as directed relation graph with error propagation (DRGEP) proposed by Pepiot-Desjardins et al.\cite{Desjardins08}, revised-DRG (r-DRG) from Luo et al.\cite{luo10} and pathway flux analysis (PFA) developed by Sun et al.\cite{sun10}. Methodological differences emerge in the calculation of the contributions or interaction coefficients between species. For example, the vanilla DRG uses the sum of absolute values of net reaction rates, while DRGEP introduces a priori error propagation, which evolving interaction coefficients to enhance the species sparsity modelling. Despite significant progresses made by DRG-based methods, two drawbacks remain. Firstly, the objective of DRG-based methods was made to indirectly reproduce the detailed chemical kinetics with reduced mechanism, by comparing the results after the calculation of ignition delay time, laminar flame speed, et al. Secondly, DRG-based methods concentrates on the reaction flux over a few main species considered rather than the reaction flux over all species as a whole, which will result in the deviation from detailed chemical kinetics.\cite{lu06, lu08} These insights suggest that the intricate global interactions among species present a considerable challenge for explicit capture of species sparsity by DRG-based methods, often resulting in suboptimal reduced mechanisms\cite{nagy09}.

It is noteworthy that the reaction sparsity can also serve as an objective for mechanism reduction. Since reactions contribute independently to chemical kinetics mathematically, the dependencies between them in the mechanism can be less complex. Many methods have been developed to exploit reaction sparsity. One such method, called detailed reduction, systematically identifies unimportant reactions by comparing reaction rates with pre-selected controlling reactions\cite{wang91}. However, identifying the controlling reaction is challenging due to the lack of a universally rigorous definition. Computational singular perturbation (CSP) has also been used to eliminate unimportant reactions by identifying them with an importance index for species\cite{Massias99, lam89, lam93, lam94, lu01, lam18, zhao19}. While CSP can produce compact reduced mechanisms at the skeletal level, it incurs high computational costs due to the need for Jacobian decomposition and mode projection, and its implementation is complex\cite{sun10}. Partial equilibrium and quasi-steady-state assumptions (QSSA) are also applied to mechanism reduction using reaction rate analysis with criteria such as small mole fractions and normalized net production rates\cite{peters85, peters87, chen88, ju94, sung98, lovs00, lu09cnf, gou10}. However, this method requires expert knowledge and a deep understanding of the mechanism and underlying chemistry, making it difficult to apply to the reduction of complex chemical reaction mechanisms\cite{lu05}. Overall, the lack of clear criteria for identifying dominant reactions without prior knowledge remains a significant obstacle for mechanism reduction strategies based on the exploitation of reaction sparsity.

In recent years, the widespread adoption of data-driven statistical learning methods in scientific data processing has been propelled by advances in computational resources. Notably, sparse statistical learning, rooted in sparse regression, emerges as a promising solution for unveiling intricate patterns in data thanks to its statistical capabilities. Concise representations are achieved through sparse constraints, such as Lasso penalty terms, enabling the faithful replication of complex structures in data with excellent compactness\cite{Tibshirani96}. This approach has demonstrated high effectiveness in extracting physical laws from high-dimensional data\cite{Brunton16, Rao23}. For instance, Rudy et al.\cite{rudy17} applied sparse statistical learning to identify control equations in a system with 50 physical variables and 256,000 physical states, showcasing its power in handling high-dimensional systems. Similarly, Harirchi et al.\cite{Harirchi20} employed sparse identification in chemical reaction systems, revealing predominant evolutionary processes within small chemical reaction networks and significantly contributing to a better understanding of such systems. However, to the best of our knowledge, the utilization of sparse learning for the reduction of complex chemical reaction mechanisms has not yet been reported.

In this work, we propose a sparse statistical learning approach for the reduction of detailed chemical reaction mechanisms. The reduced mechanism is learned to explicitly reproduce the dynamical evolution of detailed chemical kinetics, while constraining on the sparsity of the reduced reactions at the same time. A sparse weight is statistically optimized across a wide range of operating conditions to evaluate the importance of reactions. The reduced mechanism is constructed from species that participate in the identified dominant reactions. Validations are performed on the reduction of oxidation mechanisms for $n$-heptane (with 194 species) and 1,3-butadiene (with 581 species). To demonstrate the superiority of our method, we compare the results with those from traditional DRGEP methods and state-of-the-art (SOTA) methods. Under the same error limit constraints, the reduced mechanisms obtained using the sparse learning approach contain fewer species, indicating a more compact reduction of the detailed mechanisms. This advantage becomes more pronounced when reducing larger reaction mechanisms, highlighting the significant potential of sparse learning in simplifying complex chemical reaction mechanisms.

\section{Method\label{sec:method}}

\subsection{Mathematical model\label{subsec:subsec_2_1}}

The combustion reaction kinetics can be considered as a nonlinear dynamic system. The state vector $\Psi=(\psi_1,\cdots,\psi_{n_s})^\top\in\mathbb{R}^{n_s}$ represents the species concentration, where $n_s$ is the number of species in the detailed mechanism. The evolution of this state vector is governed by nonlinear ordinary differential equations
\begin{equation}
  \frac{d\Psi}{dt} = \mathbf{F}(\Psi) = \left(F_1(\Psi),\cdots,F_{n_s}(\Psi)\right)^\top \label{1}
\end{equation}
where each $F_j (\Psi)$ represents the net production rate of the $j$th species, defined by
\begin{equation}
  F_j(\Psi) = \sum_{i=1}^{n_r} v_{j,i}R_i(\Psi) \label{2}
\end{equation}
Here, the stoichiometric matrix $\mathbf{S}=[v_{j,i}]\in\mathbb{R}^{n_s\times n_r}$ and the reaction rates $\mathbf{R} = [R_i(\Psi)]\in\mathbb{R}^{n_r}$ determine the dynamics, with $n_r$ being the number of reactions. The reaction rate of the $i$th reaction is given by 
\begin{equation}
  R_i(\Psi) = k_i^+\prod_{j=1}^{n_s}\psi_j^{\alpha_{j,i}} - k_i^-\prod_{j=1}^{n_s}\psi_j^{\beta_{j,i}} \label{3}
\end{equation}
In this equation, $k_i^+$ and $k_i^-$ denote the forward and backward reaction constant of the $i$th reaction, respectively; $\alpha_{j,i}$ and $\beta_{j,i}$ denote the forward and backward stoichiometric coefficients of the $j$th species in the $i$th reaction ($\alpha_{j,i}-\beta_{j,i}=v_{j,i}$). The dynamic system can be expressed in matrix form as 
\begin{align}
  \frac{d\Psi}{dt}
  =  \begin{pmatrix} v_{1,1} & \cdots & v_{1,n_r} \\ \vdots & \ddots & \vdots \\ v_{n_s,1} & \cdots & v_{n_s,n_r} \end{pmatrix}
  \begin{pmatrix} R_1(\Psi) \\ \vdots \\ R_{n_r}(\Psi) \end{pmatrix} 
  := \mathbf{S} \mathbf{R}(\Psi) \label{equ:4}
\end{align}

\subsection{Sparse learning\label{subsec:subsec_2_2}}

The intricate nonlinear dependencies between species concentrations introduce significant complexity into modeling their interactions, posing challenges for effectively leveraging species sparsity. DRG-based methods attempt to address this complexity by explicitly modeling interactions between species with knowledge-driven formulations. However, achieving optimal formulations remains difficult. In contrast, our proposed method navigates these challenges by capitalizing on the linear relationship between elementary reaction rates and the change rates of system states, as shown in Equation (\hyperref[equ:4]{4}). In this linear relationship, reaction rates directly correspond to changes in species concentrations, facilitating a straightforward assessment of reaction dominance. Moreover, because reaction evolution relies entirely on reaction rates, these rates can be considered independent variables, unlike the complex dependencies between species that are difficult to analyze. We explore a modified system incorporating a weighted treatment of reactions within the original kinetic system  
\begin{equation}
  \begin{pmatrix} \frac{d\psi_1'}{dt} \\ \vdots \\ \frac{d\psi_{n_s}'}{dt} \end{pmatrix}
  = \mathbf{S} \left[\begin{pmatrix} w_1 \\ \vdots \\ w_{n_r} \end{pmatrix} \odot \begin{pmatrix} R_1(\Psi') \\ \vdots \\ R_{n_r}(\Psi') \end{pmatrix}\right] \label{5} \tag{5}
\end{equation}
Here, $\odot$ denotes the Hadamard product between two vectors. The objective of sparse learning is to ensure that the modified system closely matches the original system in terms of physicochemical characteristics. Practically, this means that the state changes in the modified system should closely resemble those in the original system  
\begin{equation}
  \begin{pmatrix} \frac{d\psi_1}{dt} \\ \vdots \\ \frac{d\psi_{n_s}}{dt} \end{pmatrix} \approx \begin{pmatrix} \frac{d\psi_1'}{dt} \\ \vdots \\ \frac{d\psi_{n_s}'}{dt} \end{pmatrix}
  \label{6} \tag{6}
\end{equation}
The weight vector $\mathbf{w}=(w_1,\cdots,w_{n_r})^\top\in[0,1]^{n_r}$ that achieves this objective indicates each reaction's contribution to changes in system states. When $w_i$ approaches 1, it signifies that the $i$th reaction is crucial to system evolution, whereas $w_i$ approaching 0 indicates negligible impact. The weight vector obtained through sparse learning is expected to have many components approaching zero, thereby identifying reactions with lower contributions. Consequently, this weight vector will be referred to as the {\em sparse weight} in subsequent discussions. The incorporation of sparse weight directly leverages reaction sparsity, allowing for the removal of reactions with minimal impact on system evolution. Concurrently, species participating only in these reactions are automatically eliminated, resulting in a reduced mechanism. Thus, the optimized sparse weight serves as a clear and reasonable criterion for evaluating reaction dominance. The dynamic equation described by the reduced mechanism is  
\begin{equation}
  \begin{pmatrix} \frac{d\psi_1}{dt} \\ \vdots \\ \frac{d\psi_{n_s'}}{dt} \end{pmatrix}
  = \mathbf{S} \begin{pmatrix} R_1(\Psi) \\ \vdots \\ R_{n_r'}(\Psi) \end{pmatrix} \label{7} \tag{7}
\end{equation}
where $n_s'$ and $n_r'$ denote the number of species and reactions in the reduced mechanism.

For implementation, chemical kinetic processes under various operating conditions are simulated using the detailed mechanism. The system states at all time points under these conditions serve as training samples for learning the sparse weight. Each training sample comprises the rates of change in system states and chemical reaction rates, forming a comprehensive dynamic equation for the specific state. Sparse learning is then applied to determine the sparse weight across all training samples, thoroughly exploring sparsity by considering variability across different operating conditions and system states.

\subsection{Optimization objective\label{sec:subsec_2_3}}

The dynamic equation with the introduced sparse weight can be expressed in vector form as  
\begin{equation}
  \frac{d\Psi'}{dt} = \mathbf{S}[\mathbf{w}\odot\mathbf{R}] \label{8} \tag{8}
\end{equation}
The learning objective to obtain a sparse weight that characterizes the sparsity of reactions is twofold. Firstly, the influence of the sparse weight on the system evolution should be minimal, ensuring a negligible relative deviation.
\begin{equation}
  \left|\frac{\frac{d\Psi}{dt} - \frac{d\Psi'}{dt}}{\frac{d\Psi}{dt}}\right| = 
  \left|\frac{\frac{d\Psi}{dt} - \mathbf{S}[\mathbf{w}\odot\mathbf{R}]}{\frac{d\Psi}{dt}}\right| < \varepsilon_1 \label{9} \tag{9}
\end{equation}
where $\varepsilon_1$ represents the upper limit of the relative error. Secondly, the learned sparse weight should exhibit sufficient sparsity, with many elements approaching zero, thus having a small $l_1$-norm.
\begin{equation}
  \Vert\mathbf{w}\Vert_1 = \sum_{i=1}^{n_r}w_i < \varepsilon_2 \label{10} \tag{10}
\end{equation}
where $\varepsilon_2$ imposes a sparse constraint on the weight vector. Combining these two objectives, the final optimization objective is formulated as 
\begin{equation}
  \mathcal{L} = (1-\lambda)\left\Vert\frac{\frac{d\Psi}{dt} - \mathbf{S}[\mathbf{w}\odot\mathbf{R}]}{\Vert\frac{d\Psi}{dt}\Vert_\infty}\right\Vert_2^2
  + \lambda\Vert\mathbf{w}\Vert_1 \label{11} \tag{11}
\end{equation}
The first component, known as the regression loss, is designed to ensure the prediction accuracy of the reduced mechanism. The use of the infinity norm $\left\Vert\frac{d\Psi}{dt}\right\Vert_\infty$ normalizes across different training samples, preventing numerical issues arising from zero rates of change. The second component, referred to as the sparsity loss, ensures the weight vector remains sparse. The hyperparameter $\lambda$ adjusts the preference between the two objectives during the optimization process.

In the implementation, batch gradient descent is employed to solve the optimization problem, aiming to minimize the objective function. The optimization is performed using samples $\left\{\frac{d\Psi}{dt}\in{\mathbb{R}^{N\times n_s}, \mathbf{R}\in\mathbb{R}^{N\times n_r}}\right\}$ from the training set discussed previously, where $N$ is the number of samples in one batch. \texttt{PyTorch} is utilized for its convenient automatic differentiation capabilities and GPU acceleration\cite{Paszke19}.

\section{Results\label{sec:results}}

\subsection{Detailed mechanisms and training sets\label{sec:subsec_3_1}} 

To evaluate the effectiveness of the proposed sparse learning strategy for reducing detailed chemical reaction mechanisms, we conducted experiments on the JetSurf 1.0 mechanism by Sirjean et al.\cite{Sirjean09}. This mechanism includes 194 species and 1459 reactions, providing a comprehensive description of the pyrolysis and oxidation kinetics of normal alkanes up to $n$-dodecane at high temperatures. Our study focuses on reducing the oxidation mechanism of $n$-heptane, a widely used prototypical hydrocarbon fuel in fundamental combustion experiments. The training sample set for sparse learning was derived from the zero-dimensional ignition process of the $n$-heptane/air system under various operating conditions. Each sample point captured the rates of change in system states, specifically species concentration net consumption rates and reaction rates at specific moments under particular operating conditions. The selected range of conditions included initial temperatures from 1000 K to 1600 K, initial pressures from 1 atm to 30 atm, and equivalence ratios from 0.5 to 1.5. Ignition delay time was defined as the moment during the ignition process when the temperature first exceeded 400 K above the initial temperature. The simulation duration for each condition's zero-dimensional ignition process was set to be approximately twice the ignition delay time. In total, 36 operating conditions were selected, each with 20,000 time points, resulting in a training set of 600,000 samples.

To further demonstrate the reliability of the sparse learning strategy, we investigated the reduction of larger detailed mechanisms. We selected the AramcoMech 3.0 mechanism, which consists of 581 species and 3037 reactions\cite{zhou18}. This comprehensive mechanism represents the oxidation of hydrocarbon and oxygenated $\ce{C_0}-\ce{C_4}$ fuels, including methanol, propene, 2-butene, and 1,3-butadiene. Our focus was on reducing the oxidation mechanism of 1,3-butadiene, a critical intermediate in the formation of soot and poly-aromatic hydrocarbons (PAH). The training set's operating conditions spanned initial temperatures from 800 K to 1600 K, initial pressures from 1 atm to 10 atm, and equivalence ratios from 0.5 to 1.5. Similar to the $n$-heptane study, we selected 36 operating conditions for training in sparse learning, with each condition including 20,000 time points, forming a training set of 600,000 samples. All simulations for the zero- and one-dimensional chemical kinetics with detailed and reduced mechanisms in this work were completed using the \texttt{Cantera} open-source toolkit\cite{Goodwin23}.

\subsection{Sparse learning for chemical kinetics\label{sec:subsec_3_2}} 

\begin{figure*}[t]
  \centering
  \includegraphics[width=\columnwidth]{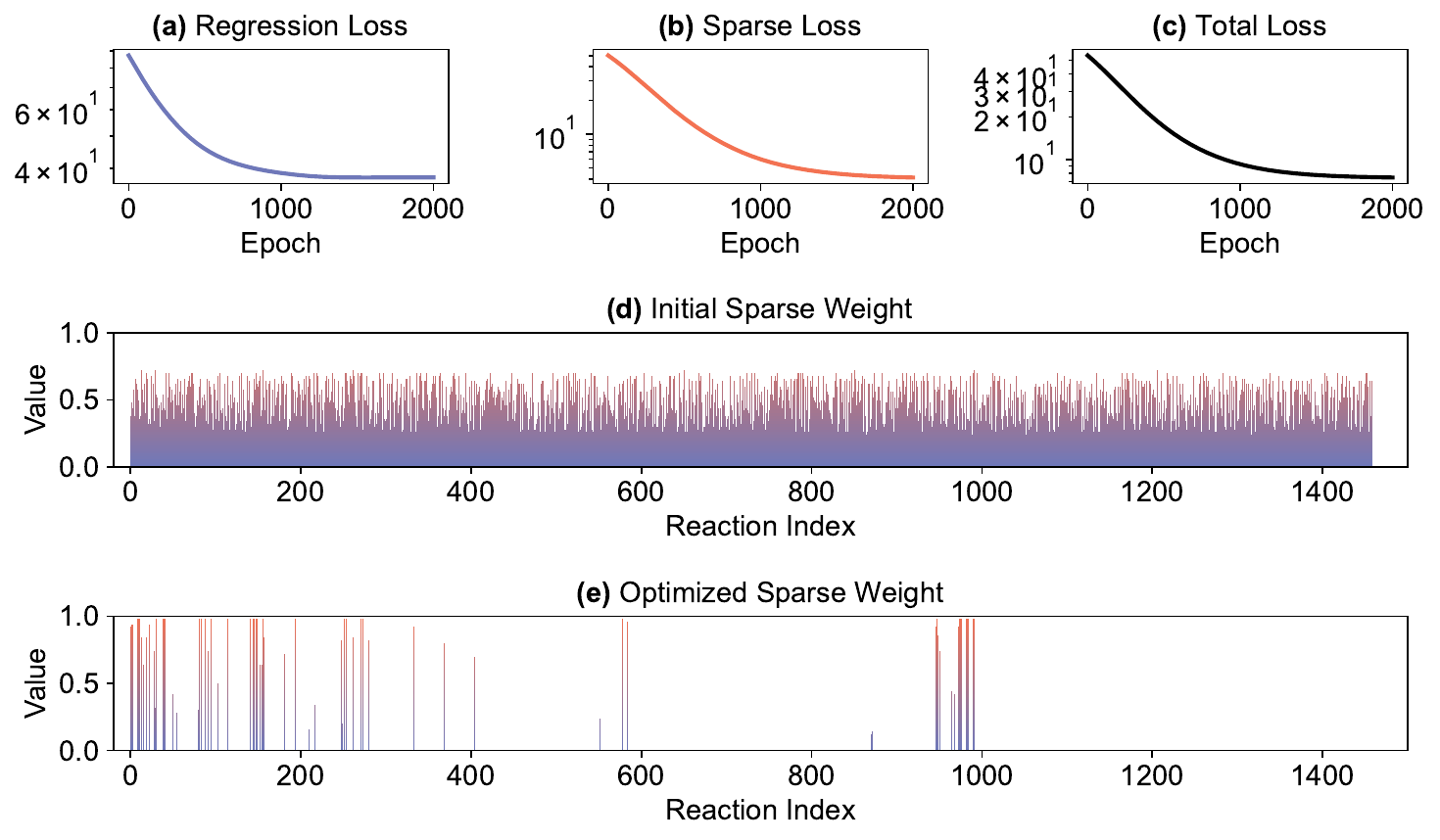}
  \caption{\footnotesize Sparse learning records for the $n$-heptane/air combustion system under the JetSurf 1.0 mechanism. (a)(b)(c) Regression, sparse and total loss profile during the optimization process. (d) The initial sparse weight distribution. (e) The optimized sparse weight distribution.}
  \label{fig:opt1}
\end{figure*}

\begin{figure*}[t]
  \centering
  \includegraphics[width=1.0\columnwidth]{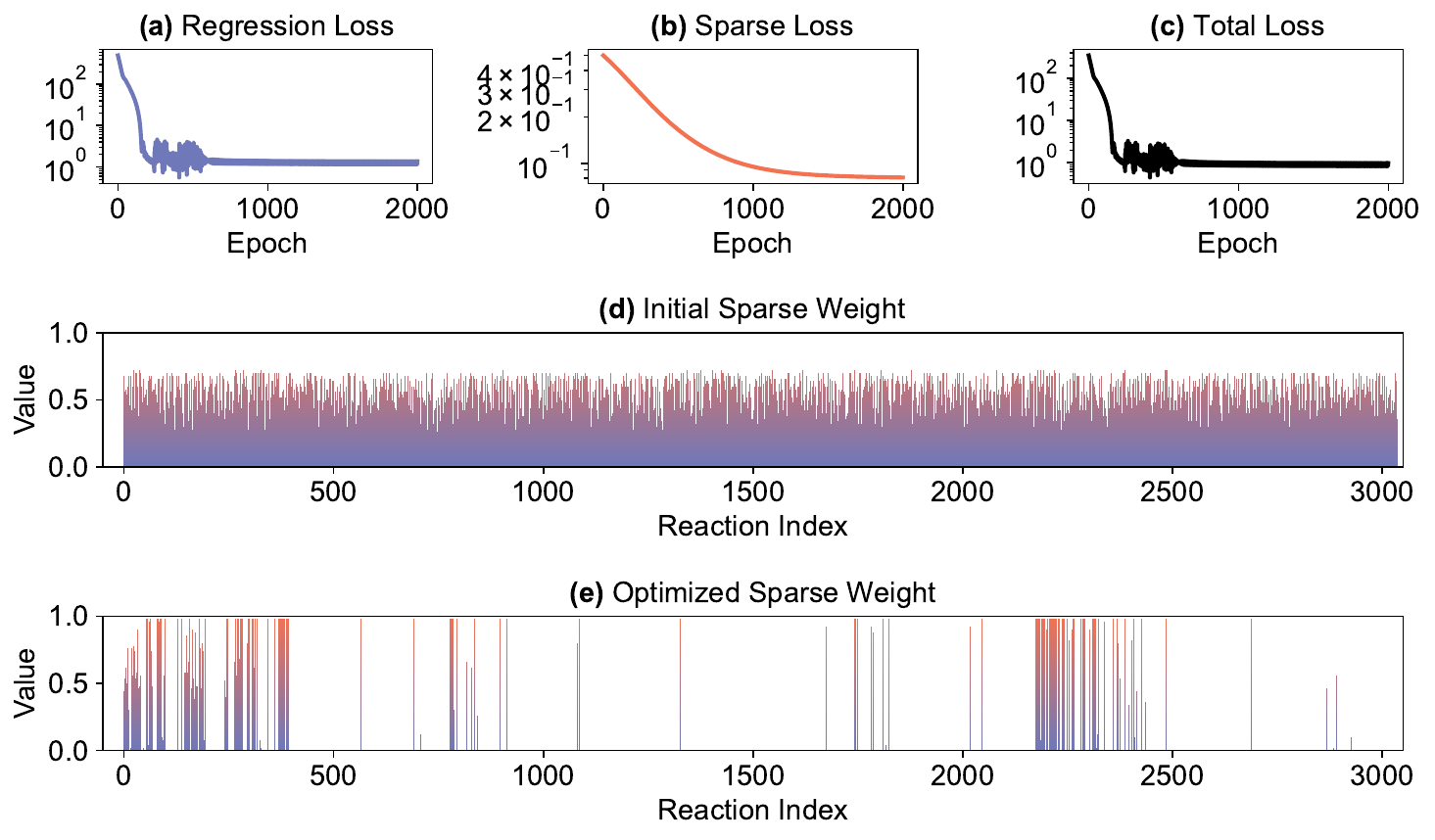}
  \caption{\footnotesize Sparse learning records for 1,3-butadiene/air combustion system under the AramcoMech 3.0 mechanism. (a)(b)(c) Regression, sparse and total loss profile during the optimization process. (d) The initial sparse weight distribution. (e) The optimized sparse weight distribution.}
  \label{fig:opt2}
\end{figure*}

Sparse learning for chemical kinetics in the detailed mechanisms was facilitated using the Adam optimizer with a batch size of 65,536, a learning rate of 5e-3, 2000 training epochs and random initialization of the weight vector in the range of 0.25 and 0.75. The sparse learning records for two different detailed mechanisms are shown in Figure \hyperref[fig:opt1]{1} and \hyperref[fig:opt2]{2}.

For the $n$-heptane/air combustion system under the JetSurf 1.0 mechanism, the hyperparameter $\lambda$ in the optimization objective is set to be 0.5. The evolution of the objective function during the training process is depicted in Figure \hyperref[fig:opt1]{1}(a),(b) and (c). The regression loss approaches convergence around epoch 1000, and the sparse loss approaches convergence around epoch 1500. The overall trend of the total loss, i.e., the objective function, aligns with that of the regression loss, indicating the overall convergence of the training process. The initial and learned weight vectors are depicted in Figure \hyperref[fig:opt1]{1}(d) and (e), respectively. The learned weight vector demonstrates noticeable variations in the magnitudes of components corresponding to different reactions. Remarkably, a limited number of weight vector components converge closely to 1 at the end of training, while the remaining components tend to approach values close to 0. This indicates a successful identification of the dominant reactions, thereby facilitating the practical acquisition of the reduced mechanism. Weight thresholds can be assigned for the selection of the reduced mechanism; reactions corresponding to sparse weight components greater than the threshold are deemed important, while those with components less than the threshold are disregarded. All species involved in these dominant reactions are collected to form the resultant reduced mechanism. The compactness of the reduced mechanism is determined by the selection of the weight threshold, with greater thresholds resulting in a more concise reduced mechanism. Reduced mechanisms with different compactness are obtained with the threshold selection. Notably, the distribution of optimized sparse weight for different hyperparameter $\lambda$ in the experiment remains the same. In other words, the relative magnitude ratio between different reaction weights remains identical, indicating that the optimization process is insensitive to the hyperparameter selection.

The optimization process and changes in sparse weights for the 1,3-butadiene/air combustion system are illustrated in Figure \hyperref[fig:opt2]{2}, where the hyperparameter $\lambda$ is set to be 0.9 for more compact reduced mechanisms. Irregular oscillations of the regression loss are observed before optimization convergence, followed by minor periodic oscillations post-convergence. These phenomena can be possibly attributed to the complexity of elementary reaction interactions in the larger reaction mechanism. A visual comparison in Figures \hyperref[fig:opt2]{2(d)} and \hyperref[fig:opt2]{2(e)} vividly reflects the sparsity of the learned weight vector. Based on the sparse weight vector, reduced mechanisms of varying scales can be obtained by selecting different weight thresholds.

\begin{figure*}[t]
  \centering
  \includegraphics[width=1.0\columnwidth]{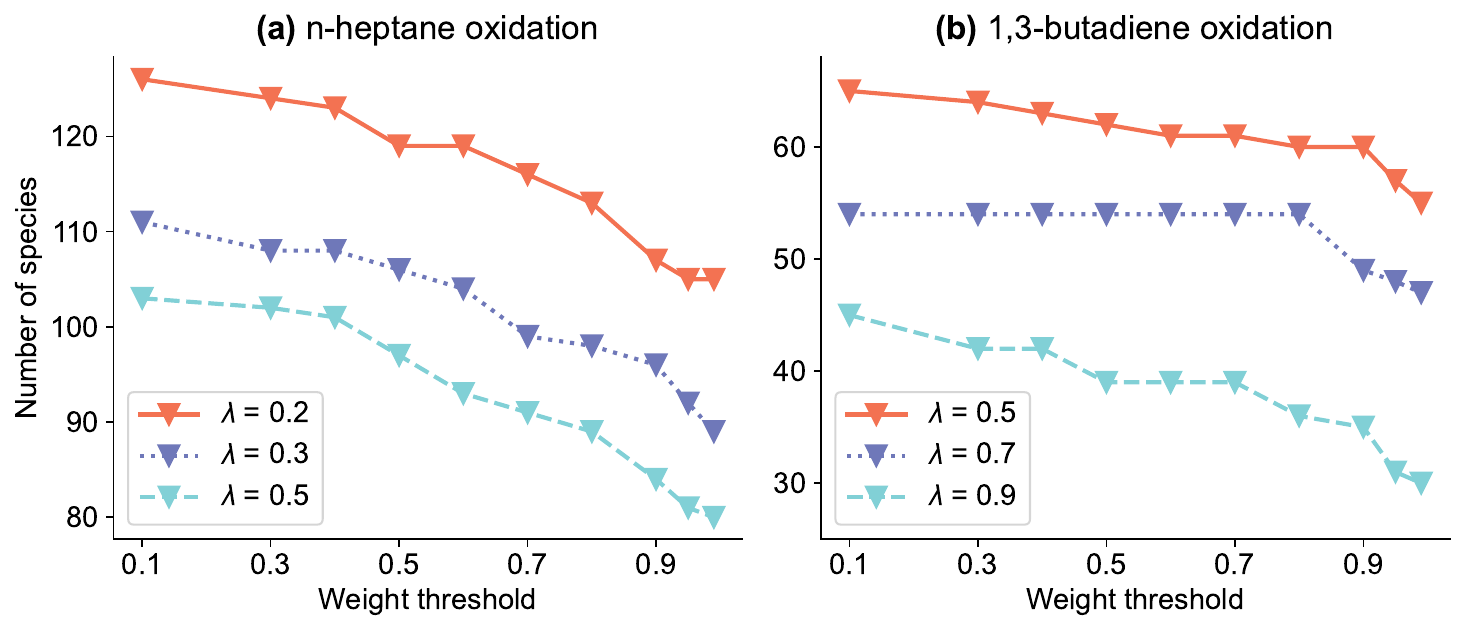}
  \caption{\footnotesize The dependence of number of species in the reduced mechanisms on the weight threshold selections in the sparse learning method for (a) $n$-heptane/air mixtures and (b) 1,3-butadiene/air mixtures. Different curves are the results under different hyperparameter $\lambda$ selections.}
  \label{fig:weight_thres}
\end{figure*}

The number of species in the reduced mechanisms, recorded under different weight threshold selections, demonstrates the progression of determining the reduced mechanisms using the sparse learning approach. These results are illustrated in Figure \hyperref[fig:weight_thres]{3}, which also compares outcomes from various hyperparameter $\lambda$ selections. It is evident that the number of species in the reduced mechanisms is a monotonically non-increasing function of the weight threshold for the same hyperparameter assignment. This indicates that larger weight thresholds intuitively yield more compact reduced mechanisms. Additionally, the number of species for a given weight threshold decreases monotonically as the hyperparameter $\lambda$ increases, with significant differences observed between the curves for different hyperparameter values. These trends highlight the importance of sparsity loss in the optimization objective, where a stronger preference for sparsity loss promotes more compact reduced mechanisms. Therefore, the selection of the hyperparameter $\lambda$ is crucial for achieving the compact yet accurate reduced chemical mechanism in the sparse learning method.

\subsection{Determination of reduced mechanisms\label{sec:subsec_3_3}}

\begin{figure*}[t]
  \centering
  \includegraphics[width=0.7\columnwidth]{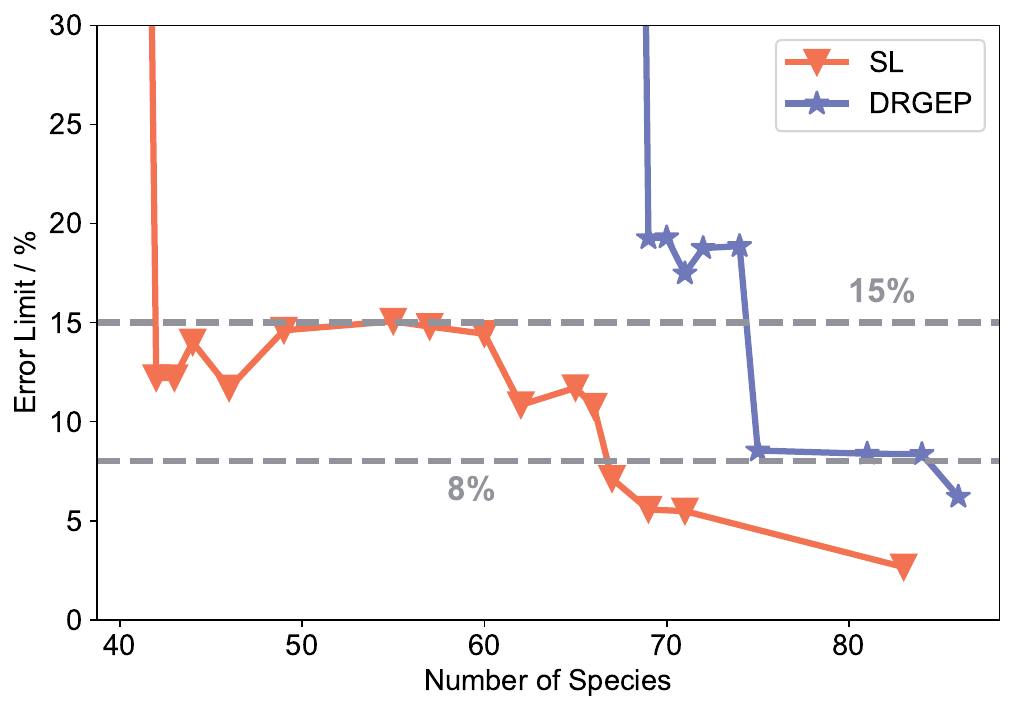}
  \caption{\footnotesize The ignition delay time error limit as a function of the number of species in the reduced mechanisms for $n$-heptane/air combustion system.}
  \label{fig:err_lim_1}
\end{figure*}

\begin{figure}[ht]
  \centering
  \includegraphics[width=0.7\columnwidth]{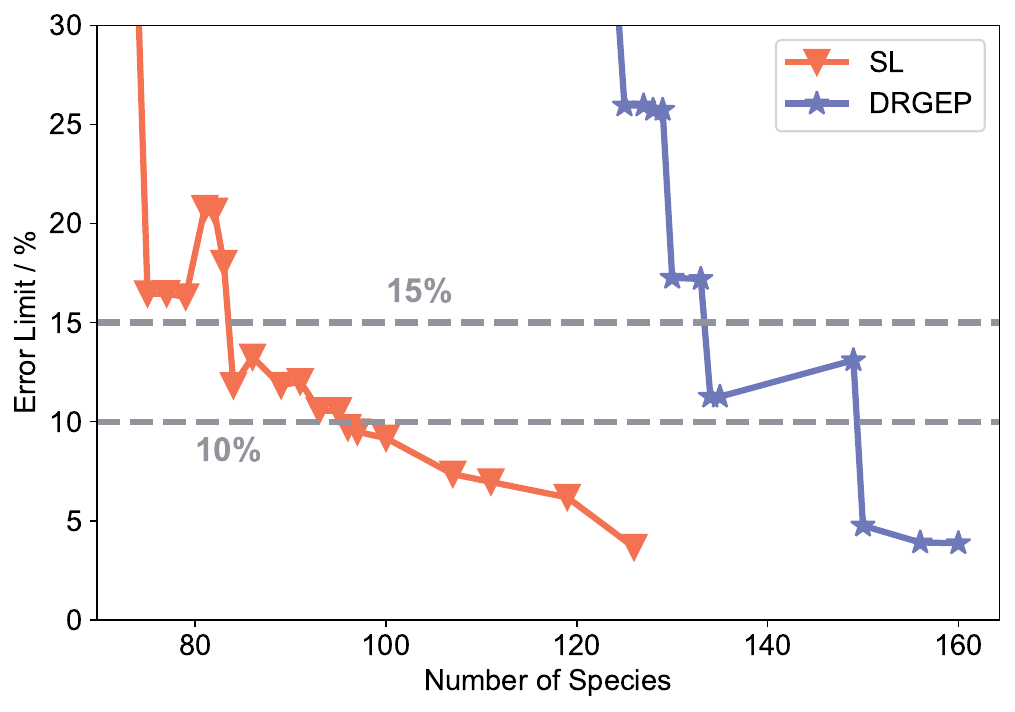}
  \caption{\footnotesize The ignition delay time error limit as a function of the number of species in the reduced mechanisms for 1,3-butadiene/air combustion system.}
  \label{fig:err_lim_2}
\end{figure}

To demonstrate the superiority of the sparse learning strategy in mechanism reduction, it was benchmarked against DRGEP method. DRGEP is renowned for its efficiency in reducing mechanisms for various hydrocarbon fuels\cite{xue20}. In this work, we utilized the DRGEP method from the \texttt{PyMARS} open-source toolkit\cite{Mestas19}, applying identical target operating conditions as those used for the sparse learning strategy. Specifically, the target species for the DRGEP method were set to be the fuels and oxidizers ($n$-heptane and oxygen for the JetSurf 1.0 mechanism, 1,3-butadiene and oxygen for the AramcoMech 3.0 mechanism), which are the default selections for DRGEP\cite{li21}.

Figure \hyperref[fig:err_lim_1]{4} illustrates the change in the ignition delay time error limit as a function of the number of species in reduced mechanisms under both the sparse learning and DRGEP methods for the JetSurf 1.0 mechanism. The results demonstrate that the sparse learning method more efficiently identifies the sparsity of species compared to the DRGEP method, resulting in simpler mechanisms for the same error limit setting. For example, at an error limit of 8\%, the DRGEP method produces a reduced mechanism with 84 species, while the sparse learning method achieves a reduced mechanism with only 67 species. Notably, when the error limit is increased to 15\%, the DRGEP method yields a mechanism with 75 species, whereas the sparse learning method results in a significantly more compact mechanism with only 43 species. In computational simulations of reactive flows, the simulation time depends linearly on the cube of the number of species for a typical implicit method, and chemical reaction calculations may consume around 90\% of the total computation time\cite{bhat03}. Therefore, the advancement of the sparse learning method over the DRGEP method is demonstrated by its ability to produce much more compact mechanisms.

The outstanding reduction capability of the sparse learning method for complex reaction mechanisms is further demonstrated by its substantial advantage in the AramcoMech 3.0 mechanism case. For the 1,3-butadiene/air combustion system, the relationship between the error limit and the number of species of reduced mechanisms obtained by both methods is shown in Figure \hyperref[fig:err_lim_2]{5}. With a maximum relative error set to 10\%, the sparse learning strategy yields a reduced mechanism with 96 species, significantly outperforming the DRGEP method, which achieves a simplified mechanism with 149 species. When the error limit is set to 15\%, the sparse learning method continues to hold a clear advantage, producing a simplified mechanism with 84 species compared to the 134 species mechanism from the DRGEP method. This advantage stems from the statistical nature of sparse learning, which provides a comprehensive understanding of the complex interactions between species and drives the profound exploration of species sparsity.

For both detailed mechanisms, we set the ignition delay time simulation error limit to 15\% to determine the reduced mechanisms. The most compact reduced mechanisms obtained using the sparse learning method and the DRGEP method are shown in Table \hyperref[tab:mechs]{1}. The table also presents the most compact reduced mechanisms obtained by state-of-the-art (SOTA) methods reported for the $n$-heptane/air combustion system under the JetSurf 1.0 mechanism\cite{lin21} and the 1,3-butadiene/air combustion system under the AramcoMech 3.0 mechanism\cite{xue20}.

\begin{table}[htbp]
\footnotesize
\renewcommand{\arraystretch}{1.2}
\centering
\caption{\footnotesize Number of species and elementary reactions of different detailed and reduced mechanisms.}
\label{tab:mechs}
\begin{tabular}{@{}cccccc@{}}
\toprule
\multicolumn{2}{c}{\multirow{2}{*}{\textbf{Methods}}} & \multicolumn{2}{c}{\textbf{JetSurf 1.0}} & \multicolumn{2}{c}{\textbf{AramcoMech 3.0}} \\ \cmidrule(l){3-6} 
\multicolumn{2}{c}{} & \begin{tabular}[c]{@{}c@{}}Number of \\ species\end{tabular} & \begin{tabular}[c]{@{}c@{}}Number of \\ elementary reactions\end{tabular} & \begin{tabular}[c]{@{}c@{}}Number of \\ species\end{tabular} & \begin{tabular}[c]{@{}c@{}}Number of \\ elementary reactions\end{tabular} \\ \midrule
\multicolumn{2}{c}{Detailed mechanism} & 194 & 1459 & 581 & 3037 \\ \midrule
\multirow{3}{*}{\begin{tabular}[c]{@{}c@{}}Reduced \\ mechanism\end{tabular}} & DRGEP & 75 & 496 & 134 & 831 \\ \cmidrule(l){2-6} 
 & \textbf{SL} & \textbf{42} & \textbf{260} & \textbf{84} & \textbf{501} \\ \cmidrule(l){2-6} 
 & SOTA & 39\textsuperscript{\cite{lin21}} & 236\textsuperscript{\cite{lin21}} & 102\textsuperscript{\cite{xue20}} & 586\textsuperscript{\cite{xue20}} \\ \bottomrule
\end{tabular}
\end{table}

\subsection{Performance of reduced mechanisms on fundamental combustion characteristics\label{sec:subsec_3_4}}

To validate the performance of reduced mechanisms obtained via the sparse learning method, comparisons with reduced mechanisms from SOTA methods were conducted. These comparisons focused on predicting fundamental combustion characteristics across a wide range of conditions for both systems.

\begin{figure*}[htbp]
  \centering
  \includegraphics[width=\columnwidth]{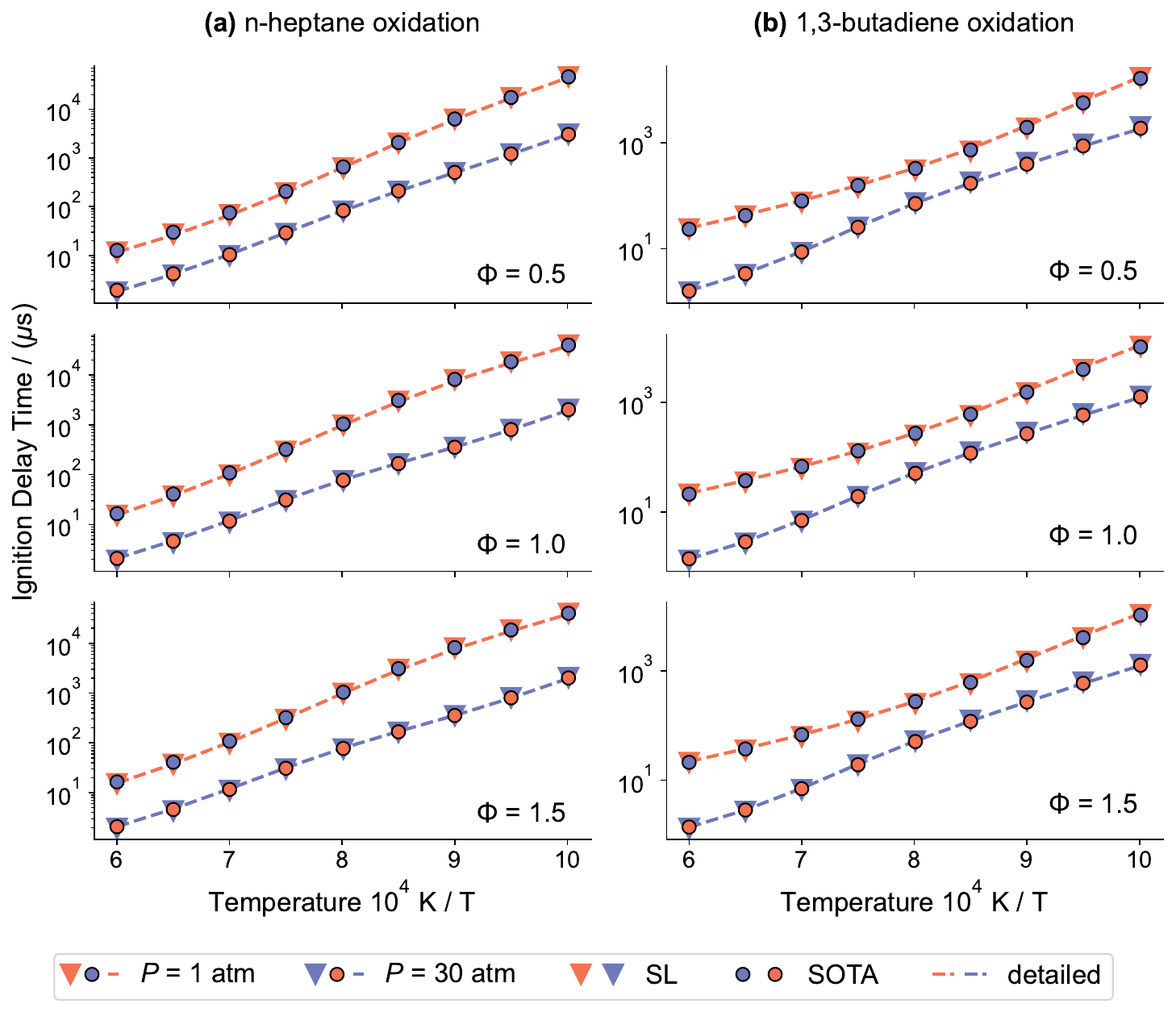}
  \caption{\footnotesize The dependence of ignition delay time on initial temperature for (a) $n$-heptane/air mixtures and (b) 1,3-butadiene/air mixtures under different initial pressure and equivalence ratio. Down triangles : reduced mechanisms from Sparse learning. Circles : reduced mechanisms from SOTA methods. Lines : detailed mechanism.}
  \label{fig:idts}
\end{figure*}

In Figure \hyperref[fig:idts]{6}, we illustrate the dependence of ignition delay time on initial temperature under different initial pressures and equivalence ratios for (a) $n$-heptane/air mixtures and (b) 1,3-butadiene/air mixtures. The results from different reaction mechanisms are represented by various lines and markers. For the $n$-heptane/air combustion system, both the sparse learning and SOTA methods demonstrated high simulation accuracy compared to the detailed mechanism. However, a comparative analysis revealed that the reduced mechanism obtained by the SOTA method, with a 14.96\% error limit, was slightly inferior to the sparse learning method, which achieved a maximum simulation relative error of 12.21\%. For the 1,3-butadiene/air combustion system, the simulation results from both reduced mechanisms also aligned well with the overall trends of the detailed mechanism. Specifically, the sparse learning reduced mechanism exhibited a maximum relative error of 11.82\%, comparable to the SOTA method's reduced mechanism, which had a maximum relative error of 8.29\%.

\begin{figure*}[htbp]
  \centering
  \includegraphics[width=\columnwidth]{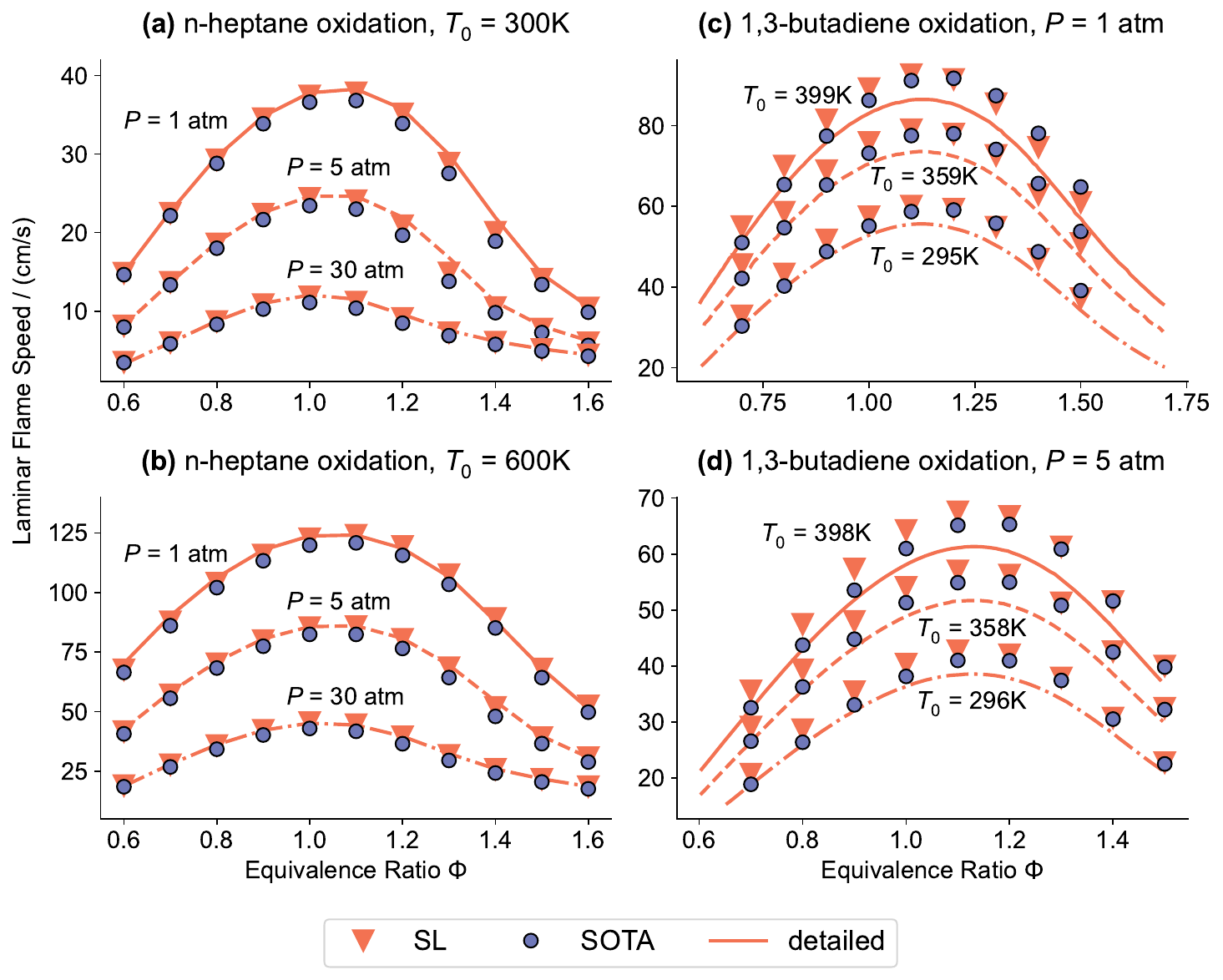}
  \caption{\footnotesize The dependence of laminar flame speed on equivalence ratio for (a)(b) $n$-heptane/air mixtures and (c)(d) 1,3-butadiene/air mixtures under different pressure and initial temperature. Down triangles : reduced mechanisms from Sparse learning. Circles : reduced mechanisms from SOTA methods. Lines : detailed mechanism.}
  \label{fig:spds}
\end{figure*}

Figure \hyperref[fig:idts]{7} shows the variation of laminar flame speed with equivalence ratio under different initial temperatures and pressures for both combustion systems. For the $n$-heptane/air combustion system, the results from the sparse learning method showed excellent consistency with the detailed mechanism, while noticeable errors were evident in the reduced mechanism from the SOTA method. Numerically, this advantage is reflected in the sparse learning method's worst-case relative error of 11.54\%, compared to the SOTA method's error limit of 17.93\%. Despite having three more species, the significant lead in error limits over the SOTA method highlights the advanced capability of the sparse learning method. For the 1,3-butadiene/air combustion system, although the results from both reduced mechanisms aligned with the general trends of the detailed mechanism, noticeable deviations were observed under every operating condition. Surprisingly, the reduced mechanism with 102 species from DRGEP had a larger error limit of 14.30\% compared to the reduced mechanism with 84 species from the sparse learning method, which achieved a maximum relative error of 11.16\%. These results indicate that the sparse learning strategy can outperform existing SOTA methods in simplifying larger reaction mechanisms, offering comparable accuracy and much more compactness.

\subsection{Performance of reduced mechanisms on turbulent combustion characteristics\label{sec:subsec_3_5}}

\begin{figure*}[htbp]
  \centering
  \includegraphics[width=\columnwidth]{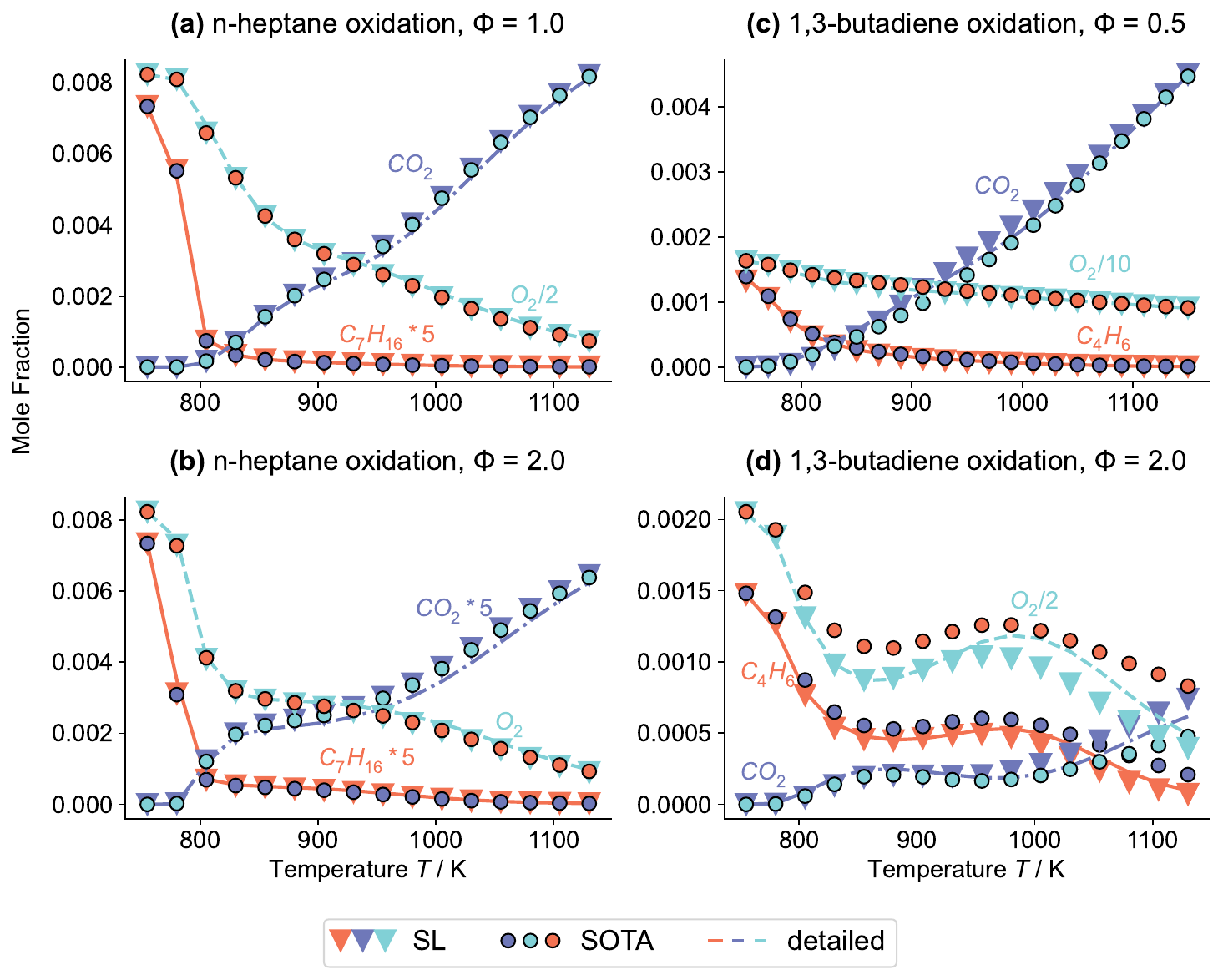}
  \caption{\footnotesize Mole fraction of different species variations with respect to temperature for (a)(b) $n$-heptane/air mixtures ($\Phi$ = 1.0 and 2.0, $P$ = 10 atm, $\tau$ = 1.0 s) and (c)(d) 1,3-butadiene/air mixtures ($\Phi$ = 0.5 and 2.0, $P$ = 10 atm, $\tau$ = 1.0 s) in perfectly stirred reactor under different operating conditions. Down triangles : reduced mechanisms from Sparse learning. Circles : reduced mechanisms from SOTA methods. Lines : detailed mechanism. ``*X" or ``/X" denotes that the plotted mole fraction has been multiplied or divided X times.}
  \label{fig:mfps}
\end{figure*}

After discussing the performance of reduced mechanisms on fundamental combustion characteristics, this subsection will analyze the predictions for turbulent combustion characteristics using partially stirred reactor (PaSR) calculations. Prior to that, the predictive performance for species mole fraction profiles was validated using perfectly stirred reactor (PSR) simulations. Figure \hyperref[fig:mfps]{8} shows the mole fractions of important species with respect to temperature under different equivalence ratios for the sparse learning, SOTA, and detailed mechanisms. For the $n$-heptane/air combustion system, good agreement is observed between the reduced and detailed mechanisms for the fuel, oxidizer, and products. These results demonstrate the reliable reproducibility of the sparse learning strategy for local chemical kinetics. In the case of the 1,3-butadiene/air combustion system, slight deviations from the detailed results appear around 1000 K at $\Phi = 2.0$ for \ce{O2} in the sparse learning mechanism. However, the SOTA mechanism shows moderate deviations for \ce{O2} across almost the entire temperature range at $\Phi = 2.0$, with a noticeable divergence trend above 1050 K. Significant departures are also observed for \ce{C4H6} and \ce{CO2} at $\Phi = 2.0$ as the temperature increases. These results highlight the superior local kinetics reproducibility of the sparse learning mechanism compared to the SOTA mechanism, despite having fewer species. Additional verifications of the sparse learning reduced mechanism for the AramcoMech 3.0 mechanism are provided in the supplementary material.

\begin{figure*}[htbp]
  \centering
  \includegraphics[width=1.0\columnwidth]{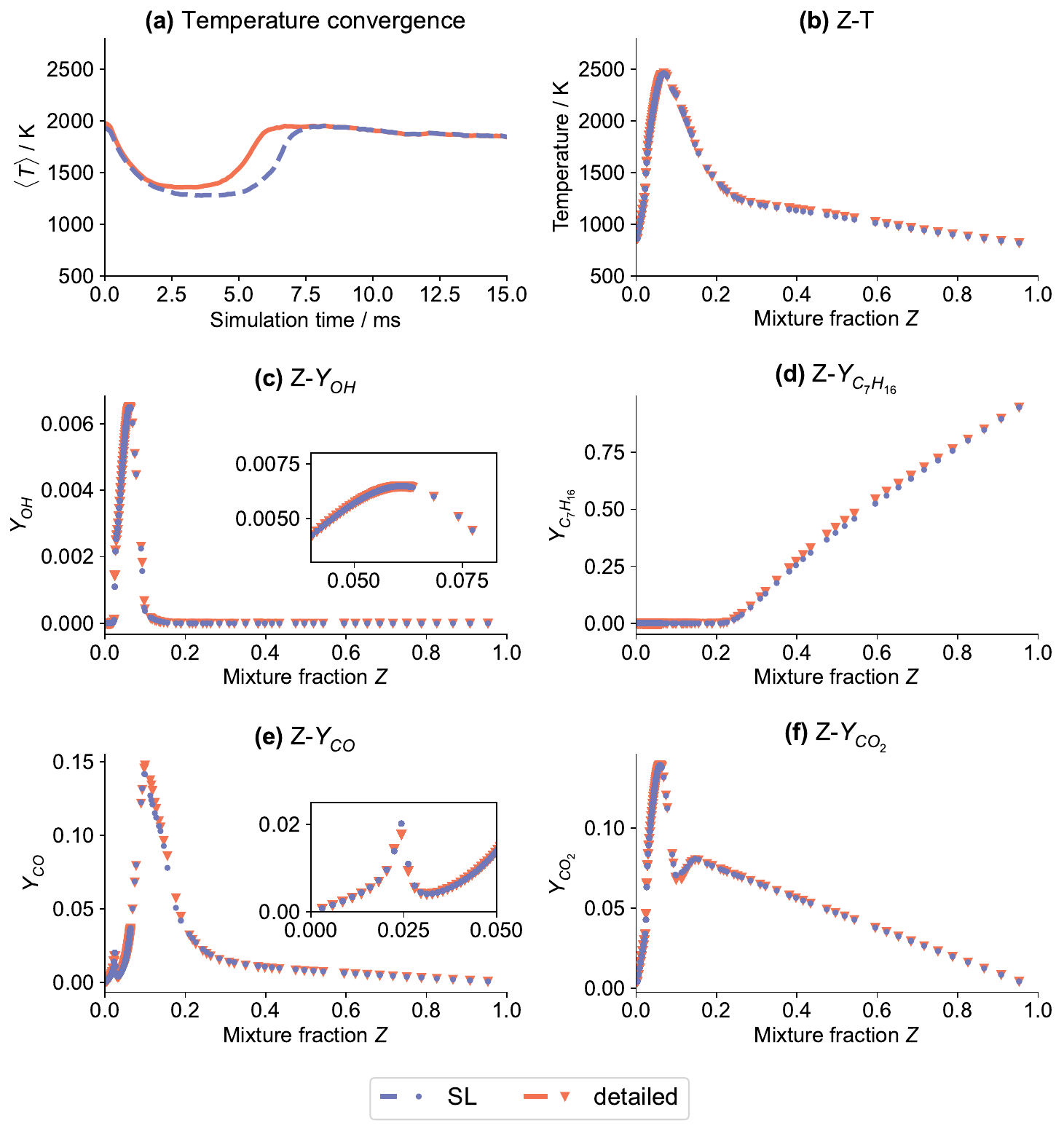}
  \caption{\footnotesize Simulation results of turbulence-chemistry interaction in partially stirred reactor for $n$-heptane/air combustion system ($T_0$ = 800 K, $P_0$ = 1.0 atm, $\Phi$ = 1.0, $\tau_\mathrm{res}$ = 3.0 ms and $\tau_\mathrm{mix}$ = 1.0 ms). Circles : reduced mechanisms from Sparse learning. Down triangles : detailed mechanism. The insets are zoom-ins of corresponding panels.}
  \label{fig:PaSR}
\end{figure*}

Next, PaSR simulations for the $n$-heptane/air combustion system were conducted to verify the accuracy of the sparse learning reduced mechanism in describing turbulence-chemistry interactions. The PaSR reactor is a powerful tool for studying small-scale turbulent mixing and combustion. Using stochastic Monte Carlo particle simulations, it can analyze the interactions between turbulent mixing and combustion\cite{yu24}. Its ability to provide useful information about the effects of mixing processes on chemical kinetics\cite{ren04} has made it a standard method for testing the capability of reduced mechanisms to simulate turbulent combustion\cite{chen97}.

In this work, the PaSR reactor parameters were set to a pressure of $P_0$ = 1.0 atm, a residence time of $\tau_\mathrm{res}$ = 3.0 ms and a mixing time of $\tau_\mathrm{mix}$ = 1.0 ms. The inlet parameters for $n$-heptane/air mixture were set to $\Phi_0$ = 1.0 and $T_0$ = 800 K, and the number of particles was set to 1000. The simulations were performed using the open-source IEM mixing model code developed by Ren's team at Tsinghua University\footnote{https://github.com/SuXY15/PaSR}. Figure \hyperref[fig:PaSR]{9} presents the simulation results of turbulence-chemistry interactions in the PaSR with $n$-heptane/air mixtures, with panels \hyperref[fig:PaSR]{9(b)-(f)} showing the distributions of temperature and key species in the mixture fraction $Z$ space. The results in Figure \hyperref[fig:PaSR]{9(a)} indicate that the simulation results have converged over the simulation time. It can be seen that the results from the sparse learning reduced mechanism are completely consistent with the detailed mechanism in terms of temperature and key species mass fractions. These results demonstrate that the sparse learning method has good reproducibility in modeling turbulence-chemistry interactions.

\section{Conclusion\label{sec:conclusion}}

In the present study, a sparse learning approach has been proposed for detailed chemical reaction mechanism reduction, treating the detailed mechanism as a dynamic system with inherent sparsity. The reaction sparsity is explicitly explored by statistically identifying the dominant reactions that control the chemical kinetics. An objective function, solvable using modern optimization methods, is established to explicitly reproduce the dynamical evolution of detailed chemical kinetics while constraining the compactness of reduced mechanisms. The performance of the sparse learning strategy was demonstrated with reductions on the JetSurf 1.0 mechanism for $n$-heptane oxidation and the AramcoMech 3.0 mechanism for 1,3-butadiene oxidation. Comparisons with existing methods validated the performance of reduced mechanisms from the sparse learning strategy in predicting both fundamental and turbulent combustion characteristics. Results showed that this approach produces reduced mechanisms with fewer species compared to the DRGEP method at the same error limit. When contrasted with SOTA methods, more thorough reductions were achieved at comparable error limits, showcasing the pioneering efficiency of sparse learning. These results are crucial, indicating that previous methods may suboptimally recognize species sparsity due to the difficulty in describing intricate species interactions. In contrast, the proposed method bypasses complex relationship modeling, profoundly exploring reaction sparsity to identify the dominant reactions with clear importance criteria optimized by sparse learning. In conclusion, the sparse learning strategy can significantly contribute to more efficient reaction mechanism reduction, thereby accelerating computational simulations of combustion reactive flows.

\section*{Declaration of competing interest}

The authors declare that they have no known competing financial interests or personal relationships that could have appeared to influence the work reported in this paper.

\section*{Acknowledgments}

This research was supported by the National Natural Science Foundation of China (grant No. U2341281, 11272026) and the High Performance Computing Center (HPCC) at Beihang University.



\bibliographystyle{elsarticle-num} 
\bibliography{sample}

\begin{thebibliography}{10}
\expandafter\ifx\csname url\endcsname\relax
  \def\url#1{\texttt{#1}}\fi
\expandafter\ifx\csname urlprefix\endcsname\relax\def\urlprefix{URL }\fi
\expandafter\ifx\csname href\endcsname\relax
  \def\href#1#2{#2} \def\path#1{#1}\fi

\bibitem{lu09}
T.~Lu, C.~K. Law, Toward accommodating realistic fuel chemistry in large-scale computations, Prog. Energy Combust. Sci. 35~(2) (2009) 192--215.

\bibitem{mass92}
U.~Maas, S.~Pope, Simplifying chemical kinetics: Intrinsic low-dimensional manifolds in composition space, Combust. Flame 88~(3) (1992) 239--264.

\bibitem{correa00}
C.~Correa, H.~Niemann, B.~Schramm, J.~Warnatz, Reaction mechanism reduction for higher hydrocarbons by the ildm method, Proc. Combust. Inst. 28~(2) (2000) 1607--1614.

\bibitem{tomlin95}
A.~S. Tomlin, M.~J. Pilling, J.~H. Merkin, J.~Brindley, N.~Burgess, A.~Gough, Reduced mechanisms for propane pyrolysis, Ind. Eng. Chem. Res. 34~(11) (1995) 3749--3760.

\bibitem{lu06}
T.~Lu, C.~K. Law, On the applicability of directed relation graphs to the reduction of reaction mechanisms, Combust. Flame 146~(3) (2006) 472--483.

\bibitem{lu05}
T.~Lu, C.~K. Law, A directed relation graph method for mechanism reduction, Proc. Combust. Inst. 30~(1) (2005) 1333--1341.

\bibitem{Desjardins08}
P.~Pepiot-Desjardins, H.~Pitsch, An efficient error-propagation-based reduction method for large chemical kinetic mechanisms, Combust. Flame 154~(1) (2008) 67--81.

\bibitem{luo10}
Z.~Luo, T.~Lu, M.~J. Maciaszek, S.~Som, D.~E. Longman, A reduced mechanism for high-temperature oxidation of biodiesel surrogates, Energy Fuels 24~(12) (2010) 6283--6293.

\bibitem{sun10}
W.~Sun, Z.~Chen, X.~Gou, Y.~Ju, A path flux analysis method for the reduction of detailed chemical kinetic mechanisms, Combust. Flame 157~(7) (2010) 1298--1307.

\bibitem{lu08}
T.~Lu, C.~K. Law, A criterion based on computational singular perturbation for the identification of quasi steady state species: A reduced mechanism for methane oxidation with no chemistry, Combust. Flame 154~(4) (2008) 761--774.

\bibitem{nagy09}
T.~Nagy, T.~Turányi, Reduction of very large reaction mechanisms using methods based on simulation error minimization, Combust. Flame 156~(2) (2009) 417--428.

\bibitem{wang91}
H.~Wang, M.~Frenklach, Detailed reduction of reaction mechanisms for flame modeling, Combust. Flame 87~(3) (1991) 365--370.

\bibitem{Massias99}
A.~Massias, D.~Diamantis, E.~Mastorakos, D.~Goussis, An algorithm for the construction of global reduced mechanisms with csp data, Combust. Flame 117~(4) (1999) 685--708.

\bibitem{lam89}
S.~Lam, D.~Goussis, Understanding complex chemical kinetics with computational singular perturbation, Symp. Int. Combust. 22~(1) (1989) 931--941.

\bibitem{lam93}
S.~Lam, Using csp to understand complex chemical kinetics, Combust. Sci. Technol. 89~(5-6) (1993) 375--404.

\bibitem{lam94}
S.~H. Lam, D.~A. Goussis, The csp method for simplifying kinetics, Int. J. Chem. Kinet. 26~(4) (1994) 461--486.

\bibitem{lu01}
T.~Lu, Y.~Ju, C.~K. Law, Complex csp for chemistry reduction and analysis, Combust. Flame 126~(1) (2001) 1445--1455.

\bibitem{lam18}
S.~Lam, An efficient implementation of computational singular perturbation, Combust. Sci. Technol. 190~(1) (2018) 157--163.

\bibitem{zhao19}
P.~Zhao, S.~Lam, Toward computational singular perturbation (csp) without eigen-decomposition, Combust. Flame 209 (2019) 63--73.

\bibitem{peters85}
N.~Peters, Numerical and asymptotic analysis of systematically reduced reaction schemes for hydrocarbon flames, in: R.~Glowinski, B.~Larrouturou, R.~Temam (Eds.), Numerical Simulation of Combustion Phenomena, Springer Berlin Heidelberg, Berlin, Heidelberg, 1985, pp. 90--109.

\bibitem{peters87}
N.~Peters, R.~Kee, The computation of stretched laminar methane-air diffusion flames using a reduced four-step mechanism, Combust. Flame 68~(1) (1987) 17--29.

\bibitem{chen88}
J.~Chen, A general procedure for constructing reduced reaction mechanisms with given independent relations, Combust. Sci. Technol. 57~(1-3) (1988) 89--94.

\bibitem{ju94}
Y.~Ju, T.~Niioka, Reduced kinetic mechanism of ignition for nonpremixed hydrogen/air in a supersonic mixing layer, Combust. Flame 99~(2) (1994) 240--246.

\bibitem{sung98}
C.~Sung, C.~Law, J.-Y. Chen, An augmented reduced mechanism for methane oxidation with comprehensive global parametric validation, Symp. Int. Combust. 27~(1) (1998) 295--304.

\bibitem{lovs00}
T.~Løvs, D.~Nilsson, F.~Mauss, Automatic reduction procedure for chemical mechanisms applied to premixed methane/air flames, Proc. Combust. Inst. 28~(2) (2000) 1809--1815.

\bibitem{lu09cnf}
T.~Lu, C.~K. Law, C.~S. Yoo, J.~H. Chen, Dynamic stiffness removal for direct numerical simulations, Combust. Flame 156~(8) (2009) 1542--1551.

\bibitem{gou10}
X.~Gou, W.~Sun, Z.~Chen, Y.~Ju, A dynamic multi-timescale method for combustion modeling with detailed and reduced chemical kinetic mechanisms, Combust. Flame 157~(6) (2010) 1111--1121.

\bibitem{Tibshirani96}
R.~Tibshirani, Regression shrinkage and selection via the lasso, J. R. Stat. Soc. B. 58~(1) (1996) 267--288.

\bibitem{Brunton16}
S.~L. Brunton, J.~L. Proctor, J.~N. Kutz, Discovering governing equations from data by sparse identification of nonlinear dynamical systems, Proc. Natl. Acad. Sci. 113~(15) (2016) 3932--3937.

\bibitem{Rao23}
C.~Rao, P.~Ren, Q.~Wang, O.~Buyukozturk, H.~Sun, Y.~Liu, Encoding physics to learn reaction--diffusion processes, Nat. Mach. Intell. 5~(7) (2023) 765--779.

\bibitem{rudy17}
S.~H. Rudy, S.~L. Brunton, J.~L. Proctor, J.~N. Kutz, Data-driven discovery of partial differential equations, Sci. Adv. 3~(4) (2017) e1602614.

\bibitem{Harirchi20}
F.~Harirchi, D.~Kim, O.~Khalil, S.~Liu, P.~Elvati, M.~Baranwal, A.~Hero, A.~Violi, On sparse identification of complex dynamical systems: A study on discovering influential reactions in chemical reaction networks, Fuel 279 (2020) 118204.

\bibitem{Paszke19}
A.~Paszke, S.~Gross, F.~Massa, A.~Lerer, J.~Bradbury, G.~Chanan, T.~Killeen, Z.~Lin, N.~Gimelshein, L.~Antiga, A.~Desmaison, A.~Kopf, E.~Yang, Z.~DeVito, M.~Raison, A.~Tejani, S.~Chilamkurthy, B.~Steiner, L.~Fang, J.~Bai, S.~Chintala, Pytorch: An imperative style, high-performance deep learning library, in: Adv. Neural Inf. Process. Syst. 32, Curran Associates, Inc., 2019, pp. 8024--8035.

\bibitem{Sirjean09}
B.~Sirjean, E.~Dames, D.~A. Sheen, X.-Q. You, C.~Sung, A.~T. Holley, F.~N. Egolfopoulos, H.~Wang, S.~S. Vasu, D.~F. Davidson, R.~K. Hanson, H.~Pitsch, C.~T. Bowman, A.~Kelley, C.~K. Law, W.~Tsang, N.~P. Cernansky, D.~L. Miller, A.~Violi, R.~P. Lindstedt, A high-temperature chemical kinetic model of n-alkane oxidation, jetsurf (Sep. 2009).

\bibitem{zhou18}
C.-W. Zhou, Y.~Li, U.~Burke, C.~Banyon, K.~P. Somers, S.~Ding, S.~Khan, J.~W. Hargis, T.~Sikes, O.~Mathieu, E.~L. Petersen, M.~AlAbbad, A.~Farooq, Y.~Pan, Y.~Zhang, Z.~Huang, J.~Lopez, Z.~Loparo, S.~S. Vasu, H.~J. Curran, An experimental and chemical kinetic modeling study of 1,3-butadiene combustion: Ignition delay time and laminar flame speed measurements, Combust. Flame 197 (2018) 423--438.

\bibitem{Goodwin23}
D.~G. Goodwin, H.~K. Moffat, I.~Schoegl, R.~L. Speth, B.~W. Weber, {Cantera}: An object-oriented software toolkit for chemical kinetics, thermodynamics, and transport processes (Aug. 2023).

\bibitem{xue20}
J.~Xue, S.~Xi, F.~Wang, An extensive study on skeletal mechanism reduction for the oxidation of c0–c4 fuels, Combust. Flame 214 (2020) 184--198.

\bibitem{Mestas19}
P.~Mestas, P.~Clayton, K.~Niemeyer, {pyMARS}: automatically reducing chemical kinetic models in {Python}, J. Open Source Softw. 4~(41) (2019) 1543.

\bibitem{li21}
H.~Li, W.~Yang, An innovative automatic dynamic target species selection technique for skeletal chemical reaction mechanism development for various fuels, Fuel 305 (2021) 121504.

\bibitem{bhat03}
B.~Bhattacharjee, D.~A. Schwer, P.~I. Barton, W.~H. Green, Optimally-reduced kinetic models: reaction elimination in large-scale kinetic mechanisms, Combust. Flame 135~(3) (2003) 191--208.

\bibitem{lin21}
S.~Lin, M.~Xie, J.~Wang, W.~Liang, C.~K. Law, W.~Zhou, B.~Yang, Chemical kinetic model reduction through species-targeted global sensitivity analysis (stgsa), Combust. Flame 224 (2021) 73--82, a dedication to Professor Ronald K. Hanson.

\bibitem{yu24}
C.~Yu, L.~Cai, J.-Y. Chen, Stochastic modeling of partially stirred reactor (pasr) for the investigation of the turbulence-chemistry interaction for the ammonia-air combustion, Flow, Turbulence and Combustion 112~(2) (2024) 509--536.

\bibitem{ren04}
Z.~Ren, S.~B. Pope, An investigation of the performance of turbulent mixing models, Combustion and Flame 136~(1-2) (2004) 208--216.

\bibitem{chen97}
J.-Y. Chen, Stochastic modeling of partially stirred reactors, Combustion Science and Technology 122~(1-6) (1997) 63--94.

\end{thebibliography}

\end{document}